%% file: main.tex
\documentclass[sigconf]{acmart}
\AtBeginDocument{%
  \providecommand\BibTeX{{%
    \normalfont B\kern-0.5em{\scshape i\kern-0.25em b}\kern-0.8em\TeX}}}

\setcopyright{acmlicensed}
\copyrightyear{2018}
\acmYear{2018}
\acmDOI{XXXXXXX.XXXXXXX}

\acmConference[Conference acronym 'XX]{Make sure to enter the correct
  conference title from your rights confirmation emai}{June 03--05,
  2018}{Woodstock, NY}
\acmISBN{978-1-4503-XXXX-X/18/06}

\usepackage{multirow}
\usepackage{threeparttable}
\usepackage{subfig}
\usepackage{pifont}
\usepackage{svg}
\usepackage[inline]{enumitem}
\begin{document}

\title{Uncovering Selective State Space Model's Capabilities in Lifelong Sequential Recommendation}

\input{definition}
\input{sections/authors}
\input{sections/abstract}

\renewcommand{\shortauthors}{Trovato and Tobin, et al.}



\maketitle

\input{sections/introduction}

\input{sections/method}
\input{sections/experiment_settings}
\input{sections/experiment_results}
\input{sections/conclusions}


\bibliographystyle{ACM-Reference-Format}
\bibliography{main}

\end{document}

%% file: definition.tex
\newcommand{\model}{Mamba4Rec}

\newcommand{\hb}[1]{{\color{red}{#1 -- HBW }}}
\newcommand{\todo}[1]{{\color{red}{#1}}}

%% file: sections/authors.tex
\author{Jiyuan Yang}
\email{jiyuan.yang@mail.sdu.edu.cn}
\affiliation{%
  \institution{Shandong University}
  \city{Qingdao}
  \country{China}
}
\author{Yuanzi Li}
\email{liyuanzi@mail.sdu.edu.cn}
\affiliation{%
  \institution{Shandong University}
  \city{Qingdao}
  \country{China}
}
\author{Jingyu Zhao}
\email{jingyu.zhao@mail.sdu.edu.cn}
\affiliation{%
  \institution{Shandong University}
  \city{Qingdao}
  \country{China}
}
\author{Hanbing Wang}
\email{hanbingwang01@gmail.com}
\affiliation{%
  \institution{Michigan State University}
  \city{East Lansing}
  \country{USA}
}


\author{Muyang Ma}
\email{muyang0331@gmail.com}
\affiliation{%
  \institution{Shandong University}
  \city{Qingdao}
  \country{China}
}

\author{Jun Ma}
\email{majun@mail.sdu.edu.cn}
\affiliation{%
  \institution{Shandong University}
  \city{Qingdao}
  \country{China}
}
\author{Zhaochun Ren}
\email{z.ren@liacs.leidenuniv.nl}
\affiliation{%
\institution{Leiden University}
\city{Leiden}
\country{The Netherlands}
}
\author{Mengqi Zhang}
\email{mengqi.zhang@sdu.edu.cn}
\affiliation{%
  \institution{Shandong University}
  \city{Qingdao}
  \country{China}
}
\author{Xin Xin}
\email{xinxin@sdu.edu.cn}
\affiliation{%
  \institution{Shandong University}
  \city{Qingdao}
  \country{China}
}
\author{Zhumin Chen}
\email{chenzhumin@sdu.edu.cn}
\affiliation{%
  \institution{Shandong University}
  \city{Qingdao}
  \country{China}
}
\author{Pengjie Ren}
\authornote{Corresponding author.}
\email{renpengjie@sdu.edu.cn}
\affiliation{%
  \institution{Shandong University}
  \city{Qingdao}
  \country{China}
}


%% file: sections/abstract.tex
\begin{abstract}
Sequential Recommenders have been widely applied in various online services, aiming to model users' dynamic interests from their sequential interactions. 
With users increasingly engaging with online platforms, vast amounts of lifelong user behavioral sequences have been generated.
However, existing sequential recommender models often struggle to handle such lifelong sequences. 
The primary challenges stem from computational complexity and the ability to capture long-range dependencies within the sequence.

Recently, a state space model featuring a selective mechanism (i.e., Mamba) has emerged. 
In this work, we investigate the performance of Mamba for lifelong sequential recommendation (i.e., length>=2k). 
More specifically, we leverage the Mamba block to model lifelong user sequences selectively. 
We conduct extensive experiments to evaluate the performance of representative sequential recommendation models in the setting of lifelong sequences. 
Experiments on two real-world datasets demonstrate the superiority of Mamba. 
We found that RecMamba achieves performance comparable to the representative model while significantly reducing training duration by approximately 70\% and memory costs by 80\%.
Codes and data are available at \url{https://github.com/nancheng58/RecMamba}.
\end{abstract}

%% file: sections/introduction.tex
\section{Introduction}

Over the past decade, recommender systems have been widely applied in various online services, e.g., online shopping~\cite{reinforce-e-commerce}, video or music platforms~\cite{nextitnet}, news recommendation~\cite{qi2020fednews}, etc., with the primary aim of offering users the most compelling items through the modeling of user interests. 
Sequential recommendation, a burgeoning field within recommendation systems, focuses on extracting dynamic user interests from their sequential interactions. The key to sequential recommendation is to capture user interests by modeling user interaction sequences.
Early studies in sequential recommendations utilize Markov Chains~\cite{DBLP:conf/www/RendleFS10,DBLP:conf/recsys/HeFWM16} for modeling user interaction sequences.
Over the past few years, plenty of deep learning-based sequential recommendation models have been proposed, including recurrent neural networks (RNN) ~\cite{DBLP:journals/corr/HidasiKBT15,DBLP:conf/recsys/HidasiQKT16}, convolutional neural networks (CNN) ~\cite{nextitnet,caser-rec}, graph neural networks (GNN) ~\cite{wu19srgnn,zhang2022dynamic},  attention-based methods ~\cite{NARM, SASRec,sun2019bert4rec} and Transformer-based methods ~\cite{de2021transformers4rec}. 

However, the aforementioned methods are constrained in their ability to model long user interaction sequences for several reasons. 
For instance, Markov Chains-based methods~\cite{DBLP:conf/www/RendleFS10,DBLP:conf/recsys/HeFWM16} assume that the next interaction is dependent ony on the previous interaction (or a few preceding ones), making it challenging to characterize long-range item transitions. 
GRU4Rec~\cite{gru4rec} may not be optimal for lifelong sequence recommendation due to information forgetting and inherent vanishing gradients. 
SASRec~\cite{SASRec} encounters challenges in modeling long sequences due to the huge memory requirements and the quadratic complexity in the input length of self-attention. 
Currently, the majority of research only allows their models to accept about 200 user interaction records~\cite{lin2023rella}.

The primary challenge in modeling lifelong user interaction sequences revolves around two key aspects: i) How to efficiently handle extended user behavior sequences, while addressing concerns like data sparsity and computational complexity? and ii) How to capture long-range item transition dependencies within the sequence? 
Recently, Mamba~\cite{gu2023mamba}, a novel state space model featuring a selective mechanism has emerged, with the aim of achieving the modeling power of Transformer~\cite{Transformer} while scaling linearly in sequence length. 
Some works leverage Mamba for various downstream tasks, such as visual tasks~\cite{zhu2024vision, liu2024vmamba}, medical applications~\cite{ma2024u}, recommendation~\cite{liu2024mamba4rec}, and so on. 
In this scenario, we are curious about examining the performance of both early models and Mamba when modeling long (i.e., length>=2k) sequences.


In this work, we investigate the performance of Mamba in sequential recommendation, especially for lifelong user behavior sequences. 
Specifically, we introduce RecMamba, a sequential recommendation framework that utilizes the Mamba block to model user preferences over time. 
By integrating the capabilities of Mamba into our framework, we aim to improve recommendation performance and better accommodate evolving user preferences in lifelong sequential recommendation scenarios. 
We conduct experiments to assess the effectiveness of prominent recommendation models for long user sequence (i.e., length>=2k) scenarios on two real-world datasets. 
We found that 
\begin{enumerate*}[leftmargin=*,nosep]
\item the performance of RecMamba improves as the length increases on two real-world datasets; and 
\item RecMamba achieves superior performance than representative sequential models. More specifically, RecMamba achieves comparable performance with the representative model SASRec while greatly reducing about 70\% training duration and 80\% memory costs.
\end{enumerate*}





%% file: sections/method.tex
\section{Leveraging Mamba for Sequential Recommendation}
State Space Models (SSMs)~\cite{gu2021efficiently, smith2022simplified} are a class of models for sequence modeling.
Recently, a novel space state model incorporating the selective mechanism (i.e., Mamba) \cite{gu2023mamba} has become a hot topic. 
Compared with prior SSMs, Mamba introduces a data-dependence selection mechanism and utilizes a parallel algorithm optimized for hardware in recurrent mode, enabling effective sequence modeling, particularly for long sequences.

Lifelong sequential recommendation is a common task involving long-sequence modeling. 
Mamba4Rec~\cite{liu2024mamba4rec} is the pioneer in utilizing Mamba for efficient sequential recommendation.
Compared to SASRec~\cite{SASRec}, the principal architectural modification in Mamba4Rec involves substituting the self-attention block with the Mamba block. 
In contrast to Mamba4Rec, our approach, Rec-Mamba, replaces the Transformer layer with the Mamba block, thereby improving the efficiency of processing lifelong sequences.

%% file: sections/experiment_settings.tex
\section{Experiments}

\subsection{\textbf{Methods for Comparison}}
To demonstrate the effectiveness of RecMamba on lifelong sequential recommendation, we compare it with a range of representative recommender models based on Transformers, RNNs, and Linear Transformer:
\begin{itemize}
\item\textbf{SASRec~\cite{SASRec}} is a method that employs the traditional self-attention block, specifically multi-head attention, to generate sequence representations.
\item\textbf{GRU4Rec~\cite{GRU}} is a pioneering method leveraging RNNs to model user action sequences for session-based recommendation, treating each user's feedback sequence as a session.
\item\textbf{Linrec~\cite{LinRec}} adopts a linear Transformer architecture that utilizes L2 norm to approximate softmax fitting.
\end{itemize}

\subsection{Datasets}
In recent years, plenty of long-term or lifelong recommendation works have emerged~\cite{huang2018improving,10.1145/3292500.3330666}. 
However, the sequence lengths discussed in these studies typically range from 10 to 200. 
We think that such sequences do not aptly represent a `lifelong' user interests. For instance, in the micro-video scenario, users watch dozens or hundreds of videos daily. 
Hence, conducting experiments assessing model performance on extended sequences is imperative.

To this end, we choose two real-world datasets with longer user interaction sequences KuaiRand\footnote{\url{https://kuairand.com/}} \cite{kuairand} and LFM-1b\footnote{\url{http://www.cp.jku.at/datasets/LFM-1b/}} \cite{schedl2016lfm} to conduct the experiments. 
All items in the sequence are arranged based on chronological order. 
We filter out users and items with less than 3 recorded interactions.

Table \ref{Datasets} summarizes the statistics of the two datasets.

\begin{table}[!t]
    \centering
    \setlength{\abovecaptionskip}{3pt}
    \begin{threeparttable}
    \caption{Dataset statistics.}
    \label{Datasets}
    \begin{tabular}{cccc}
    \toprule
    Dataset  & KuaiRand & LFM-1b\cr
    \midrule
    \#users  &27,285 & 120,322\cr
    \#items  &32,038,725 & 31,634,450\cr
    \#interactions  &322,278,385 &1,088,161,692\cr
    \#avg.len  &11,811 &9,043\cr
    \bottomrule
  \end{tabular}
    \end{threeparttable}
\end{table}

\begin{itemize}
\item \textbf{KuaiRand}. KuaiRand is a large-scale dataset derived from the famous micro-video platform Kuaishou. It uniquely features millions of randomly exposed items inserted in the standard feeds, ensuring unbiased sequential recommendations. This dataset contains 27,285 users and 32,038,725 videos and the average interaction length is 11,811.

\item \textbf{LFM-1b}. This is a large-scale dataset conducted by the music platform Last.FM, which contains more than one billion music-listening interactions covering 31,634,450 music and a total of 120,322 user clicks. The average interaction length is 9,043.
\end{itemize}

\subsection{Implementation Details}
We adopted the hyperparameter settings used for sequence recommendation due to the absence of prior work on sequence recommendation with lengths exceeding 2k. 
The trainable parameters were initialized using Xavier normal distribution. 
We employed the Adam optimizer for each model. 
For all models, we used the batch size of 256 for sequence length of 2k, while for the length of 5k, we used a batch size of 256 for RecMamba, Linrec, GRU4Rec, and a batch size of 32 for SASRec to avoid Out-of-Memory. 
The setting of the learning rate is 0.0004 and 0.0002, respectively. 
The hidden layer size is set to 50. 
We pad the sequence with an additional padding item if the sequence length is less than 2k. 
All models were trained 500 epochs on NVIDIA A800 80G. 

\subsection{Evaluation Protocols}
Following existing sequential recommendation studies \cite{SASRec, CIKM2020-S3Rec,sun2019bert4rec,wang2024rethinking}, we split data by the leave-one-out strategy. 
For an user $u$, the interaction sequence $S_u = [v_{1}^{u},v_{2}^{u},...,v_{n_u}^{u}]$, we use $([v_{1}^{u},...,v_{n_u-3}^{u}],v_{n_u-2}^{u})$ for training, $([v_{1}^{u},...,v_{n_u - 2}^{u}],v_{n_u -1}^{u})$ for validation, and $([v_{1}^{u},...,v_{n_u-1}^{u}], \\ v_{n_u}^{u})$ for testing, where $n$ is the length of $S_u$ and $v_{t}^{u}$ is the item user interact with at the $t$-th timestamp. 
We adopt Recall and NDCG metrics to evaluate the ranking performance. 
We report the average results of three experiments. 

%% file: sections/experiment_results.tex
\section{EXPERIMENTS RESULTS}
In this section, we aim to answer the following research questions in terms of lifelong sequential recommendation: 
\begin{itemize}
    \item \textbf{RQ1:} How does longer sequence length contribute to the effectiveness of sequence recommendation?
    \item \textbf{RQ2:} How do different recommenders perform in modeling lifelong sequences?
    \item \textbf{RQ3:} How is the efficiency of different recommenders?
\end{itemize}

\subsection{Lengths Comparison (RQ1)}
\begin{figure}[!tp]
    \captionsetup[subfloat]
    {}
    \subfloat[]{
        \label{a}
        \includegraphics[width=0.5\linewidth]{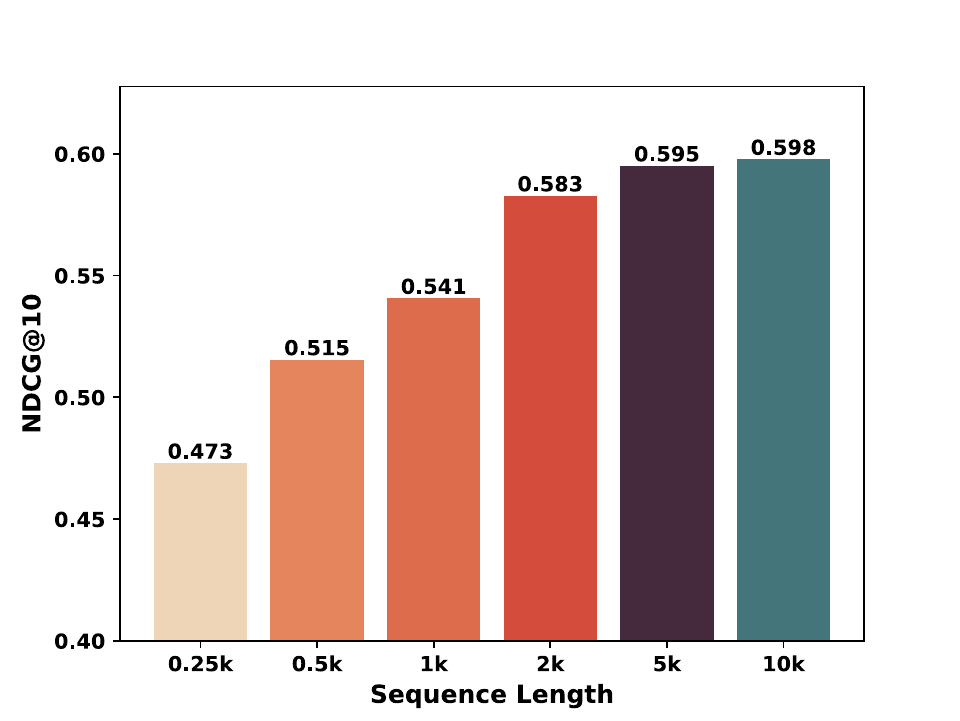}
    }
    \subfloat[]{
        \label{b}
        \includegraphics[width=0.5\linewidth]{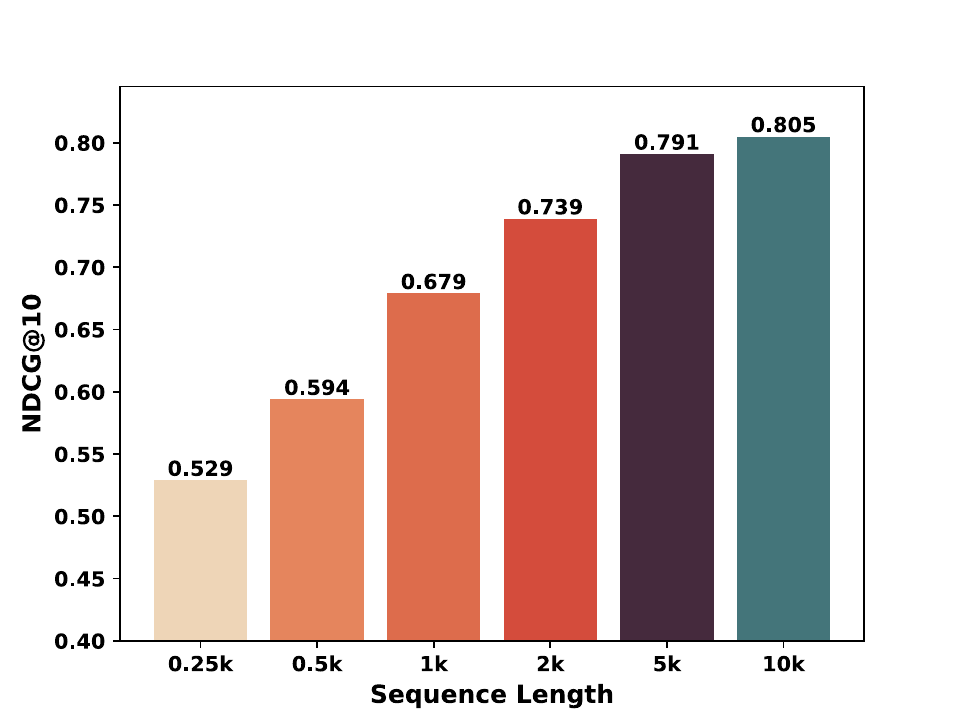}
    }
    \caption{Performance of the user interests modeling about sequence length on KuaiRand (a) and Tracks (b) datasets.}
    \label{fig:rq1}
\end{figure}
To demonstrate the performance of modeling different sequence lengths, we conduct experiments to evaluate the performance of Rec-Mamaba on two real-world datasets KuaiRand and LFM-1b. 
We report the main experimental results in Figure \ref{fig:rq1}. 

Overall, with RecMamba, we observe that longer sequences consistently outperform shorter sequences on both the KuaiRand and LFM-1b datasets, indicating that introducing longer sequences benefits modeling user interest. 
The performance improvement with longer sequences can be attributed to their ability to capture more comprehensive and meaningful patterns, dependencies, and user preferences, enabling the recommender systems to make more accurate predictions and generate more relevant recommendations. 
In summary, the experimental results provide strong evidence for the effectiveness of lifelong sequences in improving recommendation performance. 
The findings emphasize the importance of considering longer sequences in sequence recommendation models, as they can capture richer interaction information and lead to more accurate recommendations. 

\subsection{Overall Performance (RQ2)}
\label{tb:multi_result}
\begin{table*}[ht]
    \centering
    \setlength{\abovecaptionskip}{3pt}
    \begin{threeparttable}
    \caption{Overall performance comparison of different methods on KuaiRand and LFM-1b datasets. Boldface indicates the best result.
Underscore indicates the suboptimal results. 2k and 5k denote the max sequence length.}
    \renewcommand\arraystretch{1.0}
    \label{table:rq1}
    \begin{tabular}{cc|cc|cc|cc}
    \toprule
Datasets&Method&Recall@5&NDCG@5&Recall@10&NDCG@10&Recall@20&NDCG@20\cr
    \midrule
    \multirow{4}{*}{KuaiRand(2k)}
    &GRU4Rec  & 0.4062& 0.3080& 0.4439& 0.3184& 0.4617& 0.3230\cr
    &LinRec  & 0.4984& 0.3774& 0.6167  & 0.4152& 0.7242 & 0.4422\cr
    &SASRec& \textbf{0.7143}& \textbf{0.5875}&\textbf{0.7857} &\textbf{0.6126}&\textbf{0.8361} &\textbf{0.6254}\cr
    \vspace{0.04cm}
    &RecMamba& \underline{0.6907}& \underline{0.5584}& \underline{0.7652}& \underline{0.5828} & \underline{0.8172}  & \underline{0.5960}\cr
    \hline
    \multirow{4}{*}{LFM-1b(2k)}
    &GRU4Rec  & 0.3119& 0.2300&0.3597&0.2401&0.3864&0.2469\cr
    &LinRec  & 0.6163& 0.5052& 0.6955 & 0.5327& 0.7668  & 0.5501\cr
    &SASRec& \textbf{0.8112}& \textbf{0.7448}&\textbf{0.8388}&\textbf{0.7537} &\textbf{0.8598}  &\textbf{0.7591}\cr
        \vspace{0.04cm}
    &RecMamba& \underline{0.8019}& \underline{0.7487}&\underline{0.8319}&\underline{0.7393} &\underline{0.8570}  &\underline{0.7456}\cr
    \hline
    \multirow{4}{*}{KuaiRand(5k)}
    &GRU4Rec  & 0.2593& 0.2200& 0.2957 & 0.3080 & 0.2318 & 0.2350\cr
    &LinRec  & 0.5163& 0.3965& 0.6278& 0.4327 & 0.7287  & 0.4582\cr
    &SASRec& \underline{0.6422}& \underline{0.5224}&\underline{0.7370} &\underline{0.5468}&\underline{0.8042}  &\underline{0.5639}\cr
        \vspace{0.04cm}
    &RecMamba& \textbf{0.7004}& \textbf{0.5713}& \textbf{0.7733}  & \textbf{0.5951}& \textbf{0.8253} & \textbf{0.6082}\cr
    \hline
    \multirow{4}{*}{LFM-1b(5k)}
    &GRU4Rec  & 0.3009& 0.2113 & 0.3800 & 0.2366& 0.4055 & 0.2432\cr
    &LinRec  & 0.6362& 0.5296&  0.7135  & 0.5543& 0.7776 & 0.5717\cr
    &SASRec& \textbf{0.8542}& \underline{0.7820}& \textbf{0.8794}  & \underline{0.7905}& \textbf{0.8969} & \underline{0.7950}\cr
    &RecMamba& \underline{0.8485}& \textbf{0.7823}&\underline{0.8727}&\textbf{0.7908}  &\underline{0.8904} &\textbf{0.7953}\cr
    \bottomrule
    \end{tabular}
    \end{threeparttable}
\end{table*}
In this subsection, we conduct a comprehensive evaluation to verify the sequence modeling performance of RecMamba. 
We compare it with different representative recommendation models including the RNN-based model GRU4Rec, attention-based model SASRec, and linear attention-based model LinRec. 
The results are reported in Table \ref{tb:multi_result}.

We can see that RecMamba and SASRec significantly outperform both LinRec and GRU4Rec with max sequence lengths of 2k and 5k on two datasets. 
The reason may be that GRU4Rec has not fully addressed the limitations of the RNN-based architecture, such as information forgetting and intrinsic vanishing gradients issues.
Besides, the LinRec modeling sequence using approximate softmax could not fit the long-range dependency of item transition.
This highlights the superior ability of RecMamba and SASRec in the lifelong sequential recommendation. 

Compared with SASRec, RecMamba achieves suboptimal performance on sequences of length 2k, whereas it outperforms SASRec on sequences of length 5k in most cases. This indicates that RecMamba is capable of effectively modeling longer user interest sequences, potentially due to its ability to select relevant items for modeling user interests.

\subsection{Efficiency Comparison (RQ3)}
In this subsection, we conduct experiments to compare the efficiency between RecMamba and other representative sequential recommenders on the lifelong sequential recommendation. 
More specifically, we delve into crucial efficiency metrics, encompassing GPU memory consumption, training duration, and inference time. We used the batch size of 256 for sequence length of 2k for all models.
To investigate the model efficiency, we evaluate the computational cost, including  GPU memory consumption, training duration, and inference time as shown in Table \ref{tb:GPU}.

We can observe that RecMamba notably reduces GPU memory footprint and significantly slashes both inference and training times. 
Compared with SASRec on the LFM-1b(2k) dataset, RecMamba reduces about 73\% training duration, 61\% inference time, and 80\% memory costs. 
We can observe similar results on the KuaiRand(2k) dataset. 
For the two datasets of the 5k version, SASRec has encountered an out-of-memory (OOM) issue and RecMamba achieves better efficiency compared with LinRec. 
These results demonstrate RecMamba's prowess in handling lifelong sequence recommendation tasks with remarkable efficiency.

\begin{figure}
    \captionsetup[subfloat]
    {}
    \subfloat[]{
        \label{a}
        \includegraphics[width=0.48\linewidth]{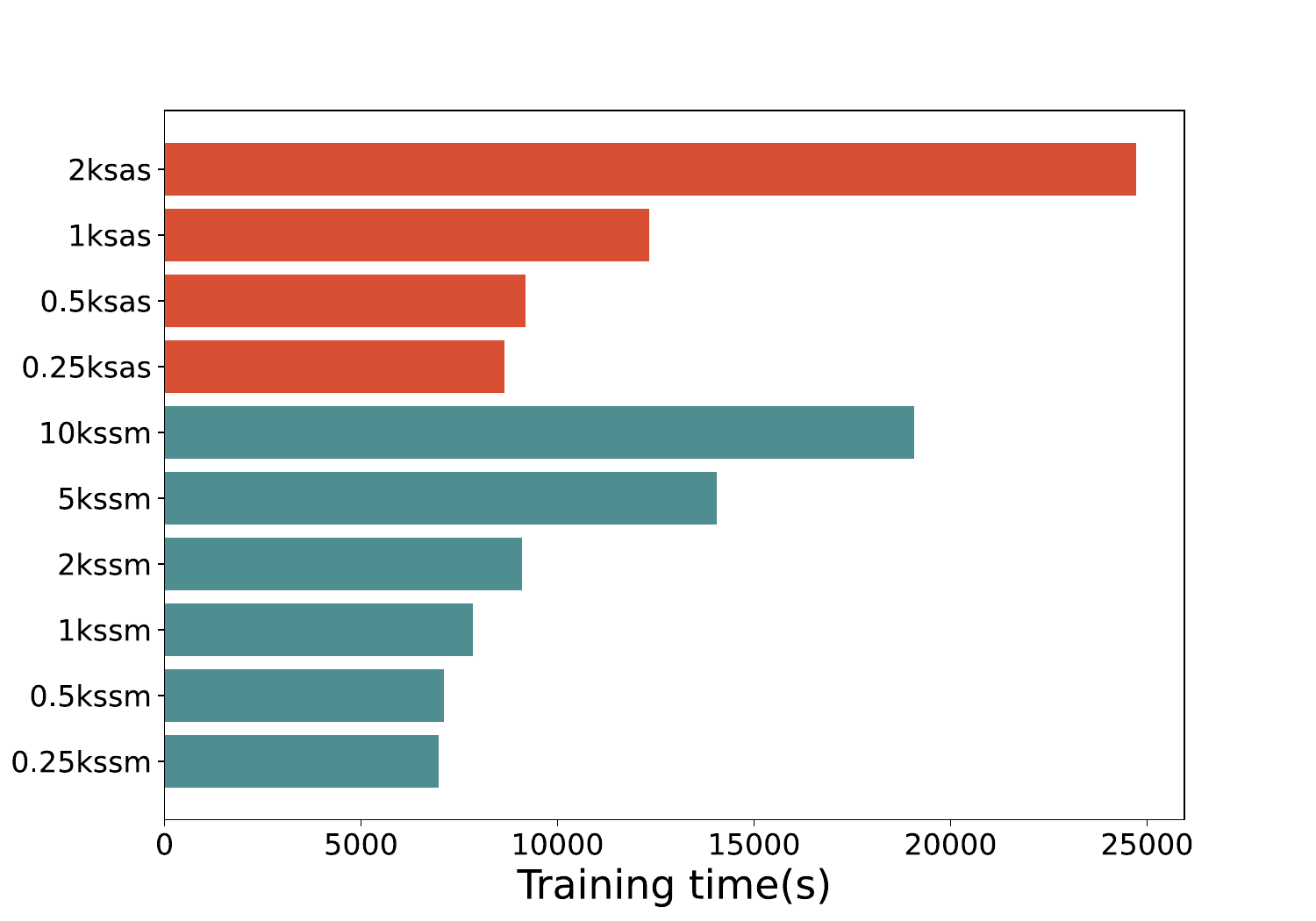}
    }
    \subfloat[]{
        \label{b}
        \includegraphics[width=0.48\linewidth]{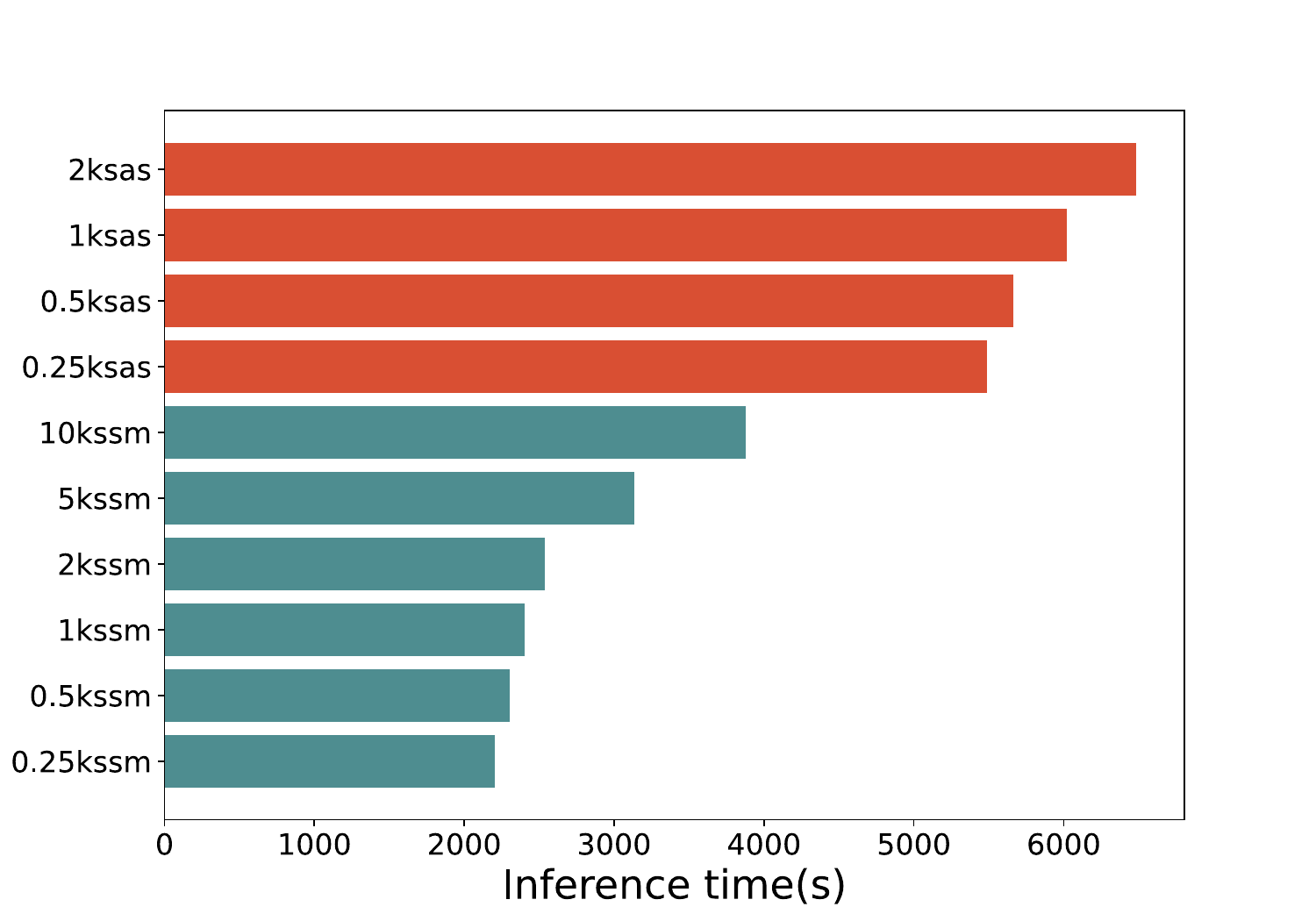}
    }
    \vspace{-0.1cm}
    \caption{Training time (a) and Inference time (b) comparision on LFM-1b dataset.}
    \vspace{-0.2cm}
    \label{fig:eff}
\end{figure}

Besides, we investigate the efficiency comparison for modeling different lengths. Figure \ref{fig:eff} shows the training time and inference time comparison between RecMamba and SASRec on the LFM-1b dataset. 
Compared with SASRec, RecMamba demonstrates a significant efficiency advantage in modeling sequences of any length reported, including both training and inference time. 
This advantage becomes more pronounced as the sequence length increases.

To conclude, the experimental findings provide compelling evidence of RecMamba's superior efficiency as a sequence recommendation framework. 
Its ability to efficiently reduce GPU memory consumption and optimize both training and inference times positions RecMamba as an exceedingly promising solution for building efficient and scalable recommendation systems.
This superiority is evident not only in terms of performance but also in terms of efficiency, including GPU memory consumption and inference time. 
This implies that RecMamba achieves better results on lifelong sequences with fewer computational resources, making it more efficient and cost-effective for deployment in recommendation systems.

\begin{table}[h]
\caption{Efficiency comparison. Boldface indicates the best result. 2k and 5k denote the max sequence length. \ding{56} denotes the OOM issue.}
\label{tb:GPU}
\centering
    \resizebox{\linewidth}{!}{
        \begin{tabular}{llccc}
    \toprule
    \textbf{Datasets} & \textbf{Model}& \textbf{GPU memory(GB)} & \multicolumn{2}{c}{\textbf{ Time cost(s/epoch)}} \\
    \cmidrule(r){4-5}
     &  &  & \textbf{Training} & \textbf{Evaluation} \\
    \midrule
                    &LinRec   &11.85G &22.49s &54.72s\\
    {KuaiRand(2k)}                                        &SASRec  &37.86G & 49.42s &64.81s\\
                                            &RecMamba  &\textbf{8.36G} &\textbf{18.21s}&\textbf{25.39s}\\
                                            \midrule
                                 &LinRec  & 20.02G & 39.54s & 61.4s\\
    KuaiRand(5K)                                        &SASRec &\ding{56} & \ding{56} & \ding{56}\\
                                            &RecMamba &\textbf{14.68G} & \textbf{28.09s} &\textbf{31.35s}\\
                                            \midrule
                                   &LinRec  &10.69G & 18.19s &52.95s\\
    LFM-1b(2k)                                        &SASRec &39.85G &41.24s &59.95s\\
                                            &RecMamba &\textbf{7.60G} &\textbf{11.75s} & \textbf{23.12s}\\
                                            \midrule
                                  &LinRec &19.83G &32.76s & 56.65s\\
    LFM-1b(5k)                                         &SASRec &\ding{56}& \ding{56} & \ding{56}\\
                                            &RecMamba &\textbf{14.46G} & \textbf{17.62s} & \textbf{26.54s}\\
    \hline
    \bottomrule
    \end{tabular}
    }
\end{table}

%% file: sections/conclusions.tex
\section{Conclusion and Future Work}
In this paper, we have investigated how the selective state space model (i.e., Mamba) performs lifelong sequential recommendation.
More specifically, we have leveraged Mamba to adopt the sequential recommendation task. 
We have conducted experiments to verify the performance and efficiency of representative recommendation models for longer user sequences (i.e., length>=2k) scenarios.
Extensive experiments and analysis on two real-world datasets have demonstrated that Mamba achieves superior performance than representative sequential models. 

In future work, we plan to refine the Mamba to better perform the subfield of recommendation. 
For instance, leveraging Mabma for multi-behavior recommendations would be a potential research direction. 
Additionally, how to effectively deal with longer side information (e.g., title and description of items) would also be an interesting research
direction. 
We hope the findings in this work could facilitate future research on using Mamba for sequential recommendations.